\documentclass[aps,prl,twocolumn,superscriptaddress]{revtex4-1}

\usepackage[]{amsmath}
\usepackage[]{graphicx}
\usepackage[caption=false]{subfig}
\usepackage[utf8]{inputenc}

\newcommand{\figref}[1]{\figurename~\ref{#1}}
\newcommand{\tabref}[1]{\tablename~\ref{#1}}

\DeclareUnicodeCharacter{0300}{}

\begin{document}

\title{Spin Current Generation by a Surface Plasmon Polariton}

\author{Theodorus Jonathan Wijaya}
\affiliation{Department of Electrical and Electronic Engineering, University of Tokyo, Hongo 7-3-1, Tokyo 113-8656, Japan}

\author{Daigo Oue}
\affiliation{Department of Physics, Imperial College London, London SW7 2BW, United Kingdom}
\affiliation{Kavli Institute for Theoretical Sciences, University of Chinese Academy of Sciences, Beijing 100190, P. R. China}

\author{Mamoru Matsuo}
\affiliation{Kavli Institute for Theoretical Sciences, University of Chinese Academy of Sciences, Beijing 100190, P. R. China}
\affiliation{CAS Center for Excellence in Topological Quantum Computation, University of Chinese Academy of Sciences, Beijing 100190, P. R. China}

\author{Yasutoshi Ito} 
\affiliation{Department of Electrical and Electronic Information Engineering, Toyohashi University of Technology, Toyohashi 441-8580, Japan}

\author{Kelvin Elphick} 
\affiliation{Department of Electronic Engineering, University of York, York YO10 5DD, United Kingdom}

\author{Hironaga Uchida} 
\affiliation{Department of Electrical and Electronic Information Engineering, Toyohashi University of Technology, Toyohashi 441-8580, Japan}

\author{Mitsuteru Inoue}
\affiliation{Department of Electrical and Electronic Information Engineering, Toyohashi University of Technology, Toyohashi 441-8580, Japan}

\author{Atsufumi Hirohata}
\affiliation{Department of Electronic Engineering, University of York, York YO10 5DD, United Kingdom}
\email{atsufumi.hirohata@york.ac.uk}

\begin{abstract}
Surface plasmon polariton (SPP) is an electromagnetic wave which is tightly localised beyond the diffraction limit at metallic surfaces. 
Recently, it is theoretically proposed that the angular momentum conversion between the SPP and electrons.
In this work, we have successfully measured SPP induced spin currents, which proves the fact that the angular momenta are interconverted. 
Such conversion from light to a spin current can be used as a coupler in a next generation spintronic computing with optical data transfer or storage.  
\end{abstract}

\maketitle

\paragraph{Introduction.---}
Spin electronics is at the verge of becoming a mainstream technology in microelectronics. 
The launch of magnetic memory production in 2018 at major microelectronic foundries marks the adoption of spintronics by microelectronics industry. 
This decisive step has now been passed and lot of further developments should yield to new applications of spintronic phenomena and devices \cite{hirohata2020review}. 
For further improvements or new directions in spintronic device applications, coupling of photonics and spintronics/nanomagnetism needs to be developed as an interdisciplinary research field in relation with the development of optical interconnects in electronics and all optical writing in storage technology.

\par
Conventionally, photoexcitation using circularly polarised light has been commonly used to generate a spin polarised electrical current mainly in a semiconductor \cite{hirohata2012optically}. 
In this method, the frequency of light needs to be comparable to the semiconductor band-gap. Thus, the method works in the limited frequency range. To exploit a wider range of frequency of light, an alternative mechanism of angular momentum conversion from optical transverse spin in surface plasmon polaritons (SPPs) to conduction electron spin has been proposed very recently \cite{oue2020electron,oue2020effects}. 

\par
Surface plasmon polaritons (SPPs) are electromagnetic waves localised at a metal/insulator interface and have been known to be excited optically with the Otto and Kretschman-Raether arrangements (or their mixture), consisting of a prism placed on a metal/insulator bilayer with and without a gap \cite{sambles1991optical}.
SPP can generate a transversely spinning electric field at the frequency of a few PHz according to the Drude-Lorentz model.
Conduction electrons in the metal can follow the SPP field \cite{bliokh2012transverse, rodriguez2013near,canaguier2014transverse,bliokh2015quantum,triolo2017spin,dai2018ultrafast,bliokh2017optical_prl,bliokh2017optical_njp}, creating their orbital motion and the corresponding inhomogeneous magnetic moments in the metal \cite{bliokh2017optical_prl,bliokh2017optical_njp}.
An earlier experimental attempt to link spintronics and photonics has been made to utilise SPP resonance to generate a spin-polarised electrical current in Au nanoparticles and films \cite{uchida2015generation}.
The spin current is induced by a localised SPP, which does not have an angular momentum.
Recently, by solving the diffusion equation under the inhomogeneous moments, Oue and Matsuo have recently predicted the generation of a spin-polarised electrical current, proposing that angular momenta are interconverted between the propagating SPP and the electronic system \cite{oue2020electron,oue2020effects}.
They estimated a spin current of $10^5\ \mathrm{A\cdot m^{-2}}$ can be generated in Au by SPP at $1.25\ \mathrm{PHz}$.
For a $20\ \mathrm{nm}$ thick Au film with a laser spot size of $1 \mathrm{mm\phi}$ the resistance can be estimated as $6.1 \times 10^{-12}\ \mathrm{\Omega}$ (resistivity of Au: $2.44 \times 10^{-8}\ \mathrm{\Omega \cdot m}$),
which can be measured using an inverse spin Hall effect.

\par
In this study, we have experimentally demonstrated the propagating SPP induced spin current by measuring the inverse spin Hall effect to prove the interconvertibility between the propagating SPP and a spin current for the first time. 
In order to confirm the presence of the SPP induced spin current, 
the other parasitic effects induced by the local heating of the laser introduction need to be removed,
such as the spin current generation from spin caloritronics. 
This has been achieved by three measurements; 
(i) reverse alignment of the inverse spin Hall effect, 
(ii) $s$- and $p$-polarisation introduction 
and 
(iii) incident angular dependence of the inverse spin Hall effect. 
The demonstrated results can be useful for the development of a SPP based opto-spintronic coupler as an interface between a spintronic device and optical data transfer or storage.

\paragraph{Experimental setup.---}
A high vacuum sputtering system (PlasmaQuest, HiTUS) was used to deposit $20\ \mathrm{nm}$ thick films of non-magnets, Ag, Pt and W.
Both a thermally oxidised Si substrate and single crystal MgO(001) substrates were used to induce SPP at the interfaces with the above metallic films.
The base pressure was $5.0 \times 10{-7}\ \mathrm{Pa}$,
while the Ar pressure for sputtering was $11\ \mathrm{Pa}$.
The films were also patterned into a Hall bar with the width of $1.2\ \mathrm{mm}$ and the length of $11\ \mathrm{mm}$,
which was designed to fit the laser spot as estimated below.

\par
\begin{figure}[htbp]
  \centering
  \includegraphics[width=\linewidth]{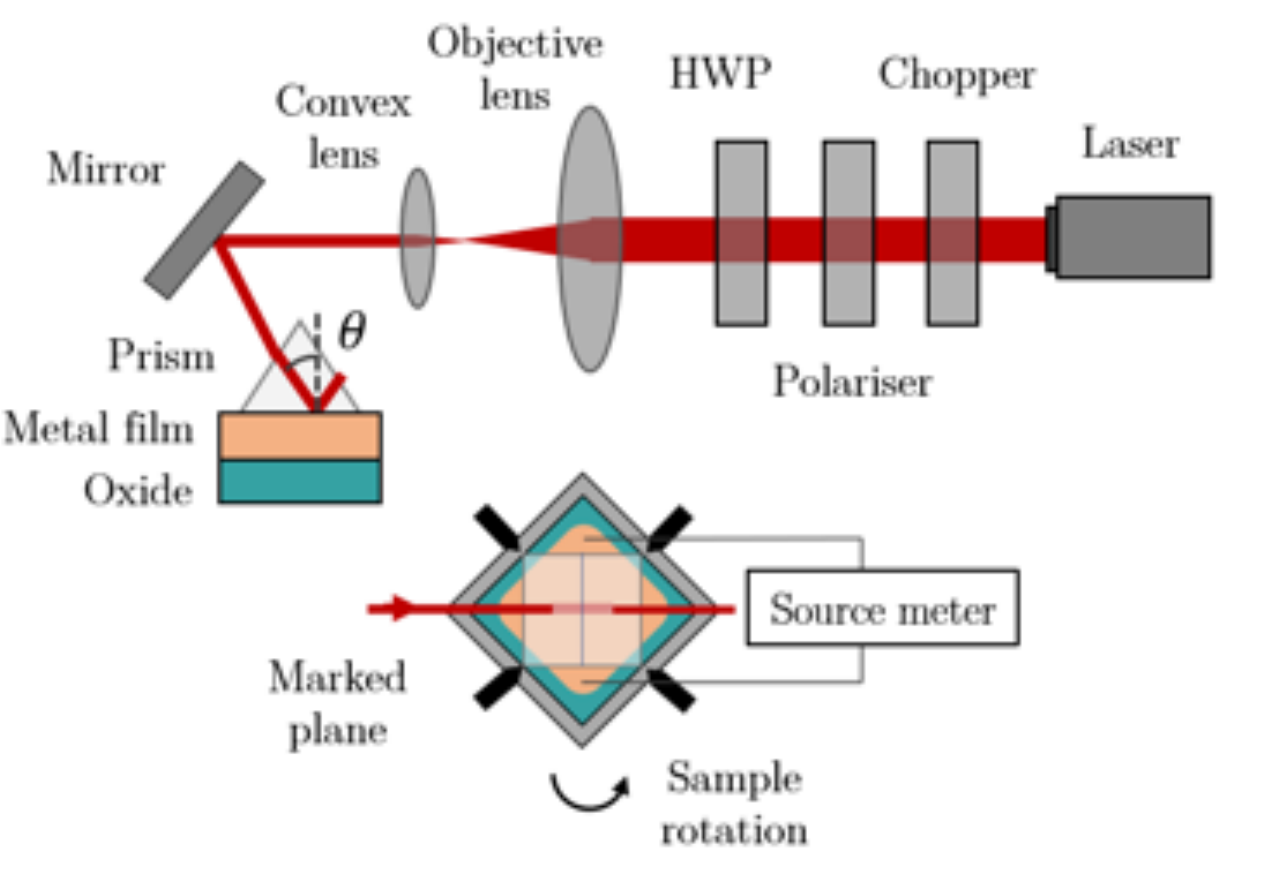}
  \caption{
    Schematic diagram of the spin current generation by SPP.
    Angle $\alpha$  indicates sample rotation with respect to its centre.
  }
  \label{fig:setup}
\end{figure}

By placing a prism (ThorLabs, N-BK7) on the surface of the Hall bar,
the Kretschmann arrangement was achieved for SPP introduction by a semiconductor laser (Thorlabs, LDM635 with the wavelength of $635\ \mathrm{nm}$ and the power of $0.4 \mathrm{mW}$).
As shown in \figref{fig:setup}, 
a linear polariser and a half-wave plate (Thorlabs, SM05 mounted zero-order $633\ \mathrm{nm}$) were used to generate both $s$- and $p$-polarisations. 
The spot size of the incident beam was then controlled using a pair of an objective lens (Mitutoyo, $2\times/0.055/f=200\ \mathrm{mm}$ M Plan) and a convex lens (Comar, Doublet 50DQ25/f=$50\ \mathrm{mm}$). 
The measured laser spot size was $1.73\ \mathrm{mm\phi}$. 
The incident angle $\theta$ of the beam to the prism was controlled by the sample stand.

\par
The generated spin current was detected using an inverse spin Hall effect as schematically shown in \figref{fig:setup}.
The source meter (Keithley, 2400) was used to detect the spin Hall voltage by connecting electrical contacts at the diagonal corners of the continuous film and the ends of the Hall bars. 
The incident angle $\theta$ was adjusted to show the largest Hall voltage.
The measurements were carried out with $1,000$ repetitions with $0.1\ \mathrm{ms}$ interval,
providing an average value with a standard deviation.
A source current was set to be $0.0$ and $0.1\ \mathrm{mA}$ for the Hall voltage and resistance measurements, respectively.

\par
The continuous films were measured to demonstrate the spin current generation by SPP as shown in \figref{fig:setup}.
Ag, Pt and W films were used for the inverse spin Hall measurement with and without a prism.
In order to subtract any parasitic effects, e.g., spin caloritronic and photovoltaic effects induced by the laser introduction
a spin current generated by the electro-motive force $I_\mathrm{emf}$ was first estimated as 
\begin{align}
  I_\textrm{emf} &= {I_\textrm{emf}}^\textrm{with prism} - {I_\textrm{emf}}^\textrm{without prism}
\end{align}
where ${I_\textrm{emf}}^\textrm{with prism}$ and ${I_\textrm{emf}}^\textrm{without prism}$ represent the current generated by the electro-motive force with and without the prism, respectively.
To offset loss caused by the prism, the transmittivity of $0.920$ 
\footnote{Thorlabs, N-BK7 \href{https://www.thorlabs.com/newgrouppage9.cfm?objectgroup_id=6973&tabname=N-BK7}{product details}} 
and the attenuation factor $\kappa = 1.2164 \times 10^{-8}$ 
\footnote{Schott, Optical glass \href{https://www.schott.com/d/advanced_optics/ac85c64c-60a0-4113-a9df-23ee1be20428/1.17/schott-optical-glass-collection-datasheets-english-may-2019.pdf}{data sheets} May 2019}
were taken into account in estimating ${j_\textrm{emf}}^\textrm{with prism}$.

\begin{figure}[htbp]
  \subfloat[\label{subfig:V_emf(a)}]{%
  \includegraphics[width=0.5\columnwidth]{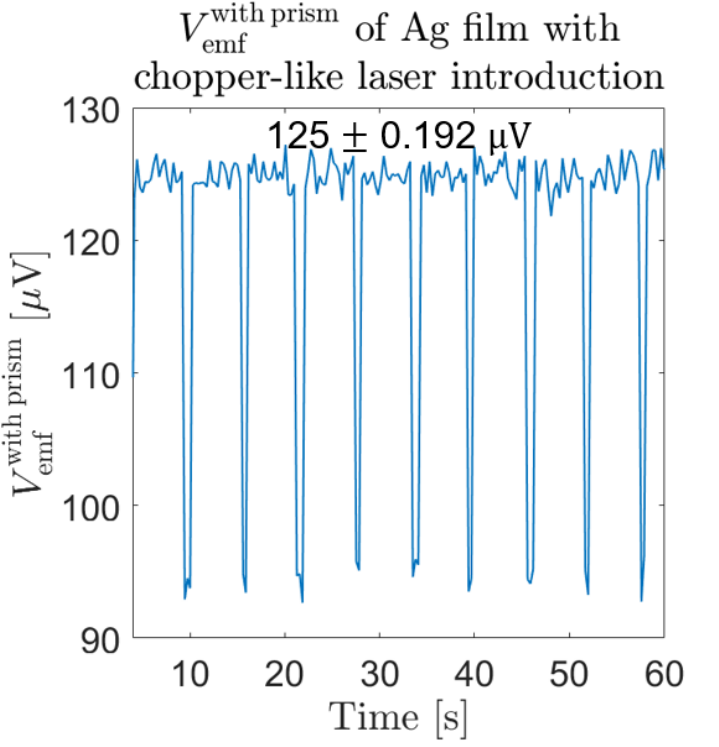}%
}\hfill
\subfloat[\label{subfig:V_emf(b)}]{%
  \includegraphics[width=0.5\columnwidth]{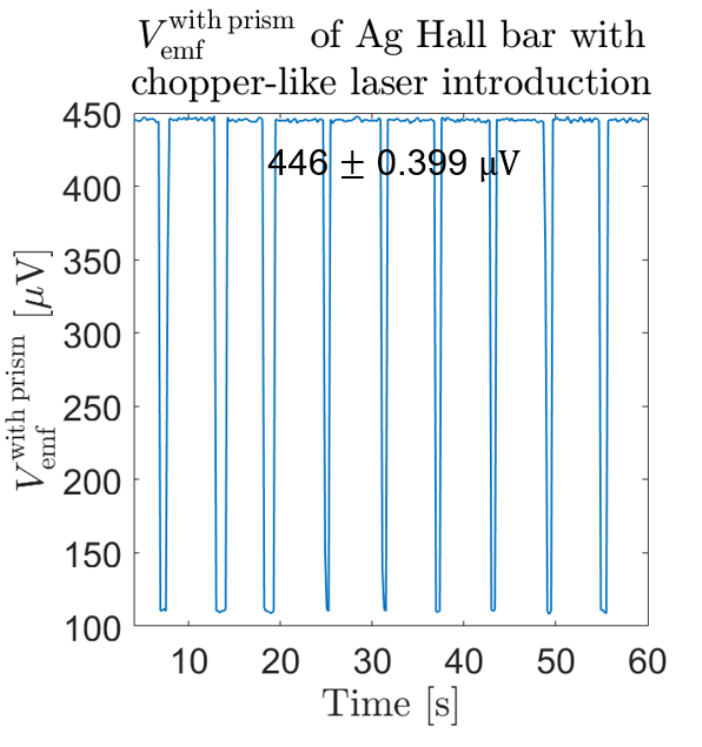}%
}
  \caption{
    Measured emf signals of chopper-like laser profile introduction to the Ag samples: 
    (a) the continuous film and (b) the Hall bar. 
    The values indicate averaged signals and standard deviations of each block of voltages under laser introduction,
    which is much larger than the background signals shown as dips.
  }
  \label{fig:V_emf}
\end{figure}

\paragraph{Experimental results.---}
The spin current generated by the electro-motive force (EMF) $I_\textrm{emf}$ was measured via a series of measurements as shown in FIGs. \ref{subfig:V_emf(a)} and \ref{subfig:V_emf(b)}, the average of which was calculated as listed in \tabref{tab:estimate}.
As clearly seen $I_\textrm{emf}$ is found to be almost proportional to $\omega$ as expected. 
By reversing the propagating direction of SPP, only the magnetic field gradient can be reversed with maintaining the thermal gradient, 
i.e., topological SPP effect by spin-momentum locking \cite{bliokh2015quantum}. 
This is confirmed by reversing the measurement configuration of the Hall voltage. 
Note that the amplitude of Iemf before and after the $180^\circ$ rotation is different but in the same order, 
which is due to the minor misalignment in the optics.

\begin{table*}[htbp]
  \centering
  \caption{
    List of estimated $I_\textrm{emf}$ in the Ag, Pt and W continuous films. 
    Individual histograms can be found in Supplemental Information.
  }
  \label{tab:estimate}
  \begin{tabular}{c|c|c|c|c}
    Samples & 
    Plasma frequency $\textrm{[PHz]}$ & 
    Damping constant $\textrm{[THz]}$ & 
    $I_\textrm{emf}\ \mathrm{[\mu A]}$ &
    $I_\textrm{emf}$ with $180^\circ$ rotation $\mathrm{[\mu A]}$
    \\
    \hline \hline
    
    Ag &
    $2.18$ \cite{ordal1985optical} &
    $4.353$ \cite{ordal1985optical} &
    $1.77 \pm 0.0453$ &
    $-2.65 \pm 0.0601$
    \\
    
    Pt &
    $1.244$ \cite{ordal1985optical} &
    $16.73$ \cite{ordal1985optical} &
    $0.619 \pm 0.00783$ &
    \\

    W &
    $0.0798 \sim 0. 967$ \cite{kffiillova1971optical} &
     &
    $0.0511 \pm 0.00476$ &

  \end{tabular}
\end{table*}

\par
Additionally, by introducing both $s$- and $p$-polarisations using a linear polariser and a half wave plate,
the Hall voltage was measured on the Ag film by the spin caloritronics (i.e., spin Nernst effect by local Joule heating) and the SPP, respectively. 
The spin current generated by SPP $I_\textrm{SPP}$ was then estimated by taking the difference between those generated by $s$- and $p$-polarisations since only the $p$-polarised light generates SPP.
\begin{align}
  I_\textrm{SPP} &={I_\textrm{emf}}^p - {I_\textrm{emf}}^s.
\end{align}
By subtracting the measured inverse spin Hall signals under the introduction of the $s$- and $p$- polarisations,
the magnitude of the SPP induced signals were estimated as listed in \tabref{tab:I_SPP}.

\begin{table}[htbp]
  \centering
  \caption{
    List of estimated Iemf and ISPP in the Ag continuous film and Hall bar. 
    Individual histograms can be found in Supplemental Information.
  }
  \label{tab:I_SPP}
  \begin{tabular}{c|c|c|c}
    Ag samples &
    ${I_\textrm{emf}}^p\ \mathrm{[\mu A]}$ &
    ${I_\textrm{emf}}^s\ \mathrm{[\mu A]}$ &
    $I_\textrm{SPP}\ \mathrm{[\mu A]}$ 
    \\
    \hline \hline

    Continuous film &
    $22.7 \pm 0.201$ &
    $20.6 \pm 0.197$ &
    $2.13 \pm 0.281$
    \\

    Hall bar &
    $19.2 \pm 0.0496$ &
    $12.6 \pm 0.162$ &
    $6.65 \pm 0.169$
  \end{tabular}
\end{table}

As a further proof of the SPP induced signals, 
incident angle dependence of these voltages was also be measured as shown in Figs. 3(a) and (b). 
The optimum incident angle to maximise emf is measured to be approximately 32.5º. 
On the other hand, by solving the wave-number matching condition between the laser and SPP,
the optimum incident angle is calculated to be as follows using the reflective indices of the prism: 
$\tilde{n} = n+i\kappa$ ($n: 1.515$ and $\kappa = 1.2164 \times 10^{-8}$) \cite{Note2}
and $\mathrm{SiO_2}: 1.457$ \cite{malitson1965interspecimen}, and the plasma frequency of Ag: $2.18\ \mathrm{PHz}$ \cite{ordal1985optical,zeman1987accurate},
\begin{align}
  \theta &= \arcsin \left( \frac{k_\textrm{SPP}(\omega_\textrm{laser})}{\omega_\textrm{laser}n_\textrm{prism}/c}\right) \approx 42.6^\circ,
\end{align}
where $k_\textrm{SPP}$ denotes the satisfying wave-number-matching condition between the wave-number of laser at the frequency $\omega_\mathrm{laser}$ and that of SPP as elaborated in Refs. \cite{maier2007plasmonics}. 
The departure may be due to several reasons in our experimental setup. 
The additional changes in the spot size of the laser at varied incident angles can change the area to generate SPP and the resulting inverse spin Hall voltages.
The fact that the rotation axis of the mirror is not on the mirror plane can also induce non-negligible errors.

\begin{figure}[htbp]
  \subfloat[\label{subfig:I_emf(a)}]{%
  \includegraphics[width=0.5\columnwidth]{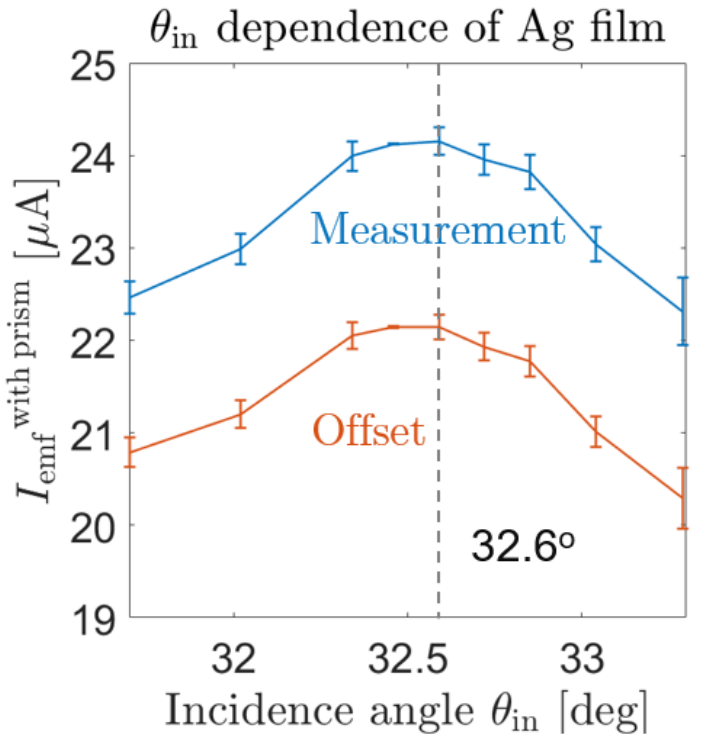}%
}\hfill
\subfloat[\label{subfig:I_emf(b)}]{%
  \includegraphics[width=0.5\columnwidth]{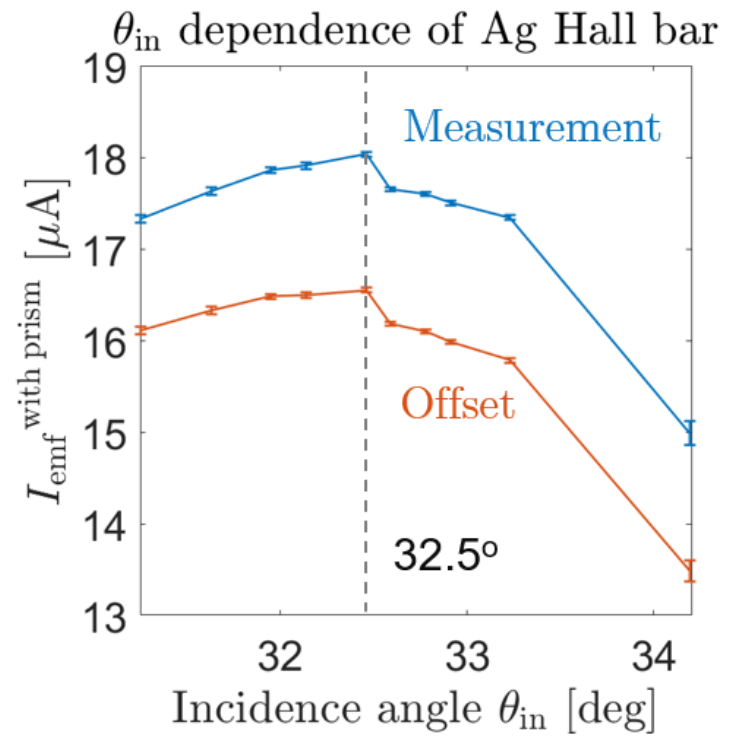}%
}
  \caption{
    Angular dependence of the EMF signals for the Ag samples:
    (a) the continuous film and (b) the Hall bar.
  }
  \label{}
\end{figure}

For the MgO(001)/Ag sample, $n$ of MgO is $1.7345$ \cite{stephens1952index}, which is larger than that of the prism, $1.515$, resulting in no SPP induced spin current to be generated. 
The measured values are $0.0612 \pm 0.0247\ \mathrm{\mu A}$,
which is almost comparable with the offset of $0.0665 \pm 0.0269\ \mu A$. 
These figures are almost fifty times smaller than those of Ag samples grown on $\mathrm{SiO_2}$.
This fact also confirms the validity of our setup to measure SPP induced spin currents.
As listed in \tabref{tab:I_SPP}, $I_\textrm{SPP}$ is estimated to be $2.128 \pm 0.281$ and $6.650 \pm 0.169\ \mathrm{\mu A}$ for the Ag continuous film and Hall bar, respectively. 
According to the theoretical calculation, 
where the SPP induced spin current ($\vec{j}_s$) can be obtained as 
\begin{align}
\vec{j}_s = \sigma_0 \nabla \delta \mu,
\end{align}
where $\sigma_0$ and $\delta \mu$ are the conductivity of the metal and the induced spin accumulation, respectively \cite{oue2020electron}.
Here $\delta \mu$ under an equilibrium state can be written as
\begin{align}
  \delta \mu = \frac{\hbar M_0}{m}
  \frac{f(\omega) (2\kappa_2 \lambda_s )^2}{(2\kappa_2 \lambda_s )^2 - 1} e^{2\kappa_2 x}.
\end{align}
Here, $M_0=(|E_0 |^2 e)/2mc$, $f(\omega)=\{2g(1-\epsilon) \sqrt{(-\epsilon)}\}/\epsilon^2$  ($g$: Gaussian unit factor and $\epsilon$: permittivity),
and $\kappa_2$: decay coefficient in the metal and $\lambda_s$: spin diffusion length in the metal. 
The upper bound of ISPP is theoretically estimated to be $100\ \mathrm{\mu A}$.
Due to the unperfect coupling from the laser beam to the SPP mode and the deviation from the surface plasmon resonance, 
the experimentally measured value can be lower than this upper bound.

\par
In our experiment, the coupling rate from the laser beam to the SPP is indeed anticipated to be lower than $100\%$, 
and the frequency of the laser is below the surface plasmon resonance frequency. By simply compare the ratio between the above theoretically estimated current and the measured one, we can estimate the conversion ratio to be $2 \sim 7\%$.
To date the selection rule governs the optoelectrical conversion especially in a direct bandgap semiconductor.
On the other hand, the conversion efficiency in the conventional semiconductor devices governed by the selection rule is limited to the spin polarisation inducible in GaAs of $\sim 40\%$ \cite{pierce1976photoemission}.
The conversion ratio can further be controlled by changing the size of the Hall bar.
A periodic array has been reported to form a wave-guide theoretically \cite{gao2018surface} and experimentally \cite{bozhevolnyi2001waveguiding},
which can be implemented in our devices to control the spin generation to maximise the efficiency.

\paragraph{Summary.---}
In summary, we have successfully measured SPP induced spin currents unambiguously with the conversion ratio of up to 7\%,
which can be further improved by optimising the device dimensions. 
Our findings prove the fact that angular momenta are interconverted between the propagating SPP and the electronic system in our setup. 
Such conversion from light to a spin current can be used as a coupler in a next generation computing with optical data transfer or storage. 
In addition, SPPs \cite{maier2007plasmonics} can be tightly localised beyond the diffraction limit at metallic surfaces and are compatible with device miniaturisation. 
The band structures of SPPs, for example, can be controlled by implementing an artificial structure in a metal or insulator thanks to the recent development in plasmonics \cite{bozhevolnyi2001waveguiding,stockman2004nanofocusing,gao2018surface}.
Such phenomena can hence revolutionise the field of optoelectronics, or rather ``opto-spintronics,'' and can open a new pathway for the spintronic devices to replace the current Si based computing devices.

\begin{acknowledgments}
 This work is financially supported by EPSRC-JSPS Core-to-Core programme (EP/M02458X/1) and JST CREST (JPMJCR17J5).
T.J.W. acknowledges the Go Global Scholarship from the University of Tokyo.
D.O. and M.M. are supported by the Priority Program of Chinese Academy of Sciences, Grant No. XDB28000000. 
Y.I. thanks the Toyohashi University of Technology for their internship programme and the financial support of the Toyoaki Scholarship Foundation.
\end{acknowledgments}


\begin{thebibliography}{26}%
\makeatletter
\providecommand \@ifxundefined [1]{%
 \@ifx{#1\undefined}
}%
\providecommand \@ifnum [1]{%
 \ifnum #1\expandafter \@firstoftwo
 \else \expandafter \@secondoftwo
 \fi
}%
\providecommand \@ifx [1]{%
 \ifx #1\expandafter \@firstoftwo
 \else \expandafter \@secondoftwo
 \fi
}%
\providecommand \natexlab [1]{#1}%
\providecommand \enquote  [1]{``#1''}%
\providecommand \bibnamefont  [1]{#1}%
\providecommand \bibfnamefont [1]{#1}%
\providecommand \citenamefont [1]{#1}%
\providecommand \href@noop [0]{\@secondoftwo}%
\providecommand \href [0]{\begingroup \@sanitize@url \@href}%
\providecommand \@href[1]{\@@startlink{#1}\@@href}%
\providecommand \@@href[1]{\endgroup#1\@@endlink}%
\providecommand \@sanitize@url [0]{\catcode `\\12\catcode `\$12\catcode
  `\&12\catcode `\#12\catcode `\^12\catcode `\_12\catcode `\%12\relax}%
\providecommand \@@startlink[1]{}%
\providecommand \@@endlink[0]{}%
\providecommand \url  [0]{\begingroup\@sanitize@url \@url }%
\providecommand \@url [1]{\endgroup\@href {#1}{\urlprefix }}%
\providecommand \urlprefix  [0]{URL }%
\providecommand \Eprint [0]{\href }%
\providecommand \doibase [0]{https://doi.org/}%
\providecommand \selectlanguage [0]{\@gobble}%
\providecommand \bibinfo  [0]{\@secondoftwo}%
\providecommand \bibfield  [0]{\@secondoftwo}%
\providecommand \translation [1]{[#1]}%
\providecommand \BibitemOpen [0]{}%
\providecommand \bibitemStop [0]{}%
\providecommand \bibitemNoStop [0]{.\EOS\space}%
\providecommand \EOS [0]{\spacefactor3000\relax}%
\providecommand \BibitemShut  [1]{\csname bibitem#1\endcsname}%
\let\auto@bib@innerbib\@empty
\bibitem [{\citenamefont {Hirohata}\ \emph {et~al.}(2020)\citenamefont
  {Hirohata}, \citenamefont {Yamada}, \citenamefont {Nakatani}, \citenamefont
  {Prejbeanu}, \citenamefont {Di{\'e}ny}, \citenamefont {Pirro},\ and\
  \citenamefont {Hillebrands}}]{hirohata2020review}%
  \BibitemOpen
  \bibfield  {author} {\bibinfo {author} {\bibfnamefont {A.}~\bibnamefont
  {Hirohata}}, \bibinfo {author} {\bibfnamefont {K.}~\bibnamefont {Yamada}},
  \bibinfo {author} {\bibfnamefont {Y.}~\bibnamefont {Nakatani}}, \bibinfo
  {author} {\bibfnamefont {L.}~\bibnamefont {Prejbeanu}}, \bibinfo {author}
  {\bibfnamefont {B.}~\bibnamefont {Di{\'e}ny}}, \bibinfo {author}
  {\bibfnamefont {P.}~\bibnamefont {Pirro}},\ and\ \bibinfo {author}
  {\bibfnamefont {B.}~\bibnamefont {Hillebrands}},\ }\bibfield  {title}
  {\bibinfo {title} {Review on spintronics: Principles and device
  applications},\ }\href@noop {} {\bibfield  {journal} {\bibinfo  {journal}
  {Journal of Magnetism and Magnetic Materials}\ }\textbf {\bibinfo {volume}
  {509}},\ \bibinfo {pages} {166711} (\bibinfo {year} {2020})}\BibitemShut
  {NoStop}%
\bibitem [{\citenamefont {Hirohata}\ and\ \citenamefont
  {Kim}(2012)}]{hirohata2012optically}%
  \BibitemOpen
  \bibfield  {author} {\bibinfo {author} {\bibfnamefont {A.}~\bibnamefont
  {Hirohata}}\ and\ \bibinfo {author} {\bibfnamefont {J.}~\bibnamefont {Kim}},\
  }\href@noop {} {\emph {\bibinfo {title} {Optically induced and detected spin
  current}}}\ (\bibinfo  {publisher} {London, UK: Oxford Univ. Press},\
  \bibinfo {year} {2012})\BibitemShut {NoStop}%
\bibitem [{\citenamefont {Oue}\ and\ \citenamefont
  {Matsuo}(2020{\natexlab{a}})}]{oue2020electron}%
  \BibitemOpen
  \bibfield  {author} {\bibinfo {author} {\bibfnamefont {D.}~\bibnamefont
  {Oue}}\ and\ \bibinfo {author} {\bibfnamefont {M.}~\bibnamefont {Matsuo}},\
  }\bibfield  {title} {\bibinfo {title} {Electron spin transport driven by
  surface plasmon polaritons},\ }\href@noop {} {\bibfield  {journal} {\bibinfo
  {journal} {Physical Review B}\ }\textbf {\bibinfo {volume} {101}},\ \bibinfo
  {pages} {161404} (\bibinfo {year} {2020}{\natexlab{a}})}\BibitemShut
  {NoStop}%
\bibitem [{\citenamefont {Oue}\ and\ \citenamefont
  {Matsuo}(2020{\natexlab{b}})}]{oue2020effects}%
  \BibitemOpen
  \bibfield  {author} {\bibinfo {author} {\bibfnamefont {D.}~\bibnamefont
  {Oue}}\ and\ \bibinfo {author} {\bibfnamefont {M.}~\bibnamefont {Matsuo}},\
  }\bibfield  {title} {\bibinfo {title} {Effects of surface plasmons on spin
  currents in a thin film system},\ }\href@noop {} {\bibfield  {journal}
  {\bibinfo  {journal} {New Journal of Physics}\ }\textbf {\bibinfo {volume}
  {22}},\ \bibinfo {pages} {033040} (\bibinfo {year}
  {2020}{\natexlab{b}})}\BibitemShut {NoStop}%
\bibitem [{\citenamefont {Sambles}\ \emph {et~al.}(1991)\citenamefont
  {Sambles}, \citenamefont {Bradbery},\ and\ \citenamefont
  {Yang}}]{sambles1991optical}%
  \BibitemOpen
  \bibfield  {author} {\bibinfo {author} {\bibfnamefont {J.}~\bibnamefont
  {Sambles}}, \bibinfo {author} {\bibfnamefont {G.}~\bibnamefont {Bradbery}},\
  and\ \bibinfo {author} {\bibfnamefont {F.}~\bibnamefont {Yang}},\ }\bibfield
  {title} {\bibinfo {title} {Optical excitation of surface plasmons: an
  introduction},\ }\href@noop {} {\bibfield  {journal} {\bibinfo  {journal}
  {Contemporary physics}\ }\textbf {\bibinfo {volume} {32}},\ \bibinfo {pages}
  {173} (\bibinfo {year} {1991})}\BibitemShut {NoStop}%
\bibitem [{\citenamefont {Bliokh}\ and\ \citenamefont
  {Nori}(2012)}]{bliokh2012transverse}%
  \BibitemOpen
  \bibfield  {author} {\bibinfo {author} {\bibfnamefont {K.~Y.}\ \bibnamefont
  {Bliokh}}\ and\ \bibinfo {author} {\bibfnamefont {F.}~\bibnamefont {Nori}},\
  }\bibfield  {title} {\bibinfo {title} {Transverse spin of a surface
  polariton},\ }\href@noop {} {\bibfield  {journal} {\bibinfo  {journal}
  {Physical Review A}\ }\textbf {\bibinfo {volume} {85}},\ \bibinfo {pages}
  {061801} (\bibinfo {year} {2012})}\BibitemShut {NoStop}%
\bibitem [{\citenamefont {Rodr{\'\i}guez-Fortu{\~n}o}\ \emph
  {et~al.}(2013)\citenamefont {Rodr{\'\i}guez-Fortu{\~n}o}, \citenamefont
  {Marino}, \citenamefont {Ginzburg}, \citenamefont {O'Connor}, \citenamefont
  {Mart{\'\i}nez}, \citenamefont {Wurtz},\ and\ \citenamefont
  {Zayats}}]{rodriguez2013near}%
  \BibitemOpen
  \bibfield  {author} {\bibinfo {author} {\bibfnamefont {F.~J.}\ \bibnamefont
  {Rodr{\'\i}guez-Fortu{\~n}o}}, \bibinfo {author} {\bibfnamefont
  {G.}~\bibnamefont {Marino}}, \bibinfo {author} {\bibfnamefont
  {P.}~\bibnamefont {Ginzburg}}, \bibinfo {author} {\bibfnamefont
  {D.}~\bibnamefont {O'Connor}}, \bibinfo {author} {\bibfnamefont
  {A.}~\bibnamefont {Mart{\'\i}nez}}, \bibinfo {author} {\bibfnamefont {G.~A.}\
  \bibnamefont {Wurtz}},\ and\ \bibinfo {author} {\bibfnamefont {A.~V.}\
  \bibnamefont {Zayats}},\ }\bibfield  {title} {\bibinfo {title} {Near-field
  interference for the unidirectional excitation of electromagnetic guided
  modes},\ }\href@noop {} {\bibfield  {journal} {\bibinfo  {journal} {Science}\
  }\textbf {\bibinfo {volume} {340}},\ \bibinfo {pages} {328} (\bibinfo {year}
  {2013})}\BibitemShut {NoStop}%
\bibitem [{\citenamefont {Canaguier-Durand}\ and\ \citenamefont
  {Genet}(2014)}]{canaguier2014transverse}%
  \BibitemOpen
  \bibfield  {author} {\bibinfo {author} {\bibfnamefont {A.}~\bibnamefont
  {Canaguier-Durand}}\ and\ \bibinfo {author} {\bibfnamefont {C.}~\bibnamefont
  {Genet}},\ }\bibfield  {title} {\bibinfo {title} {Transverse spinning of a
  sphere in a plasmonic field},\ }\href@noop {} {\bibfield  {journal} {\bibinfo
   {journal} {Physical Review A}\ }\textbf {\bibinfo {volume} {89}},\ \bibinfo
  {pages} {033841} (\bibinfo {year} {2014})}\BibitemShut {NoStop}%
\bibitem [{\citenamefont {Bliokh}\ \emph {et~al.}(2015)\citenamefont {Bliokh},
  \citenamefont {Smirnova},\ and\ \citenamefont {Nori}}]{bliokh2015quantum}%
  \BibitemOpen
  \bibfield  {author} {\bibinfo {author} {\bibfnamefont {K.~Y.}\ \bibnamefont
  {Bliokh}}, \bibinfo {author} {\bibfnamefont {D.}~\bibnamefont {Smirnova}},\
  and\ \bibinfo {author} {\bibfnamefont {F.}~\bibnamefont {Nori}},\ }\bibfield
  {title} {\bibinfo {title} {Quantum spin hall effect of light},\ }\href@noop
  {} {\bibfield  {journal} {\bibinfo  {journal} {Science}\ }\textbf {\bibinfo
  {volume} {348}},\ \bibinfo {pages} {1448} (\bibinfo {year}
  {2015})}\BibitemShut {NoStop}%
\bibitem [{\citenamefont {Triolo}\ \emph {et~al.}(2017)\citenamefont {Triolo},
  \citenamefont {Cacciola}, \citenamefont {Patan{\`e}̀}, \citenamefont
  {Saija}, \citenamefont {Savasta},\ and\ \citenamefont
  {Nori}}]{triolo2017spin}%
  \BibitemOpen
  \bibfield  {author} {\bibinfo {author} {\bibfnamefont {C.}~\bibnamefont
  {Triolo}}, \bibinfo {author} {\bibfnamefont {A.}~\bibnamefont {Cacciola}},
  \bibinfo {author} {\bibfnamefont {S.}~\bibnamefont {Patan{\`e}̀}}, \bibinfo
  {author} {\bibfnamefont {R.}~\bibnamefont {Saija}}, \bibinfo {author}
  {\bibfnamefont {S.}~\bibnamefont {Savasta}},\ and\ \bibinfo {author}
  {\bibfnamefont {F.}~\bibnamefont {Nori}},\ }\bibfield  {title} {\bibinfo
  {title} {Spin-momentum locking in the near field of metal nanoparticles},\
  }\href@noop {} {\bibfield  {journal} {\bibinfo  {journal} {ACS Photonics}\
  }\textbf {\bibinfo {volume} {4}},\ \bibinfo {pages} {2242} (\bibinfo {year}
  {2017})}\BibitemShut {NoStop}%
\bibitem [{\citenamefont {Dai}\ \emph {et~al.}(2018)\citenamefont {Dai},
  \citenamefont {Dabrowski}, \citenamefont {Apkarian},\ and\ \citenamefont
  {Petek}}]{dai2018ultrafast}%
  \BibitemOpen
  \bibfield  {author} {\bibinfo {author} {\bibfnamefont {Y.}~\bibnamefont
  {Dai}}, \bibinfo {author} {\bibfnamefont {M.}~\bibnamefont {Dabrowski}},
  \bibinfo {author} {\bibfnamefont {V.~A.}\ \bibnamefont {Apkarian}},\ and\
  \bibinfo {author} {\bibfnamefont {H.}~\bibnamefont {Petek}},\ }\bibfield
  {title} {\bibinfo {title} {Ultrafast microscopy of spin-momentum-locked
  surface plasmon polaritons},\ }\href@noop {} {\bibfield  {journal} {\bibinfo
  {journal} {ACS nano}\ }\textbf {\bibinfo {volume} {12}},\ \bibinfo {pages}
  {6588} (\bibinfo {year} {2018})}\BibitemShut {NoStop}%
\bibitem [{\citenamefont {Bliokh}\ \emph
  {et~al.}(2017{\natexlab{a}})\citenamefont {Bliokh}, \citenamefont
  {Bekshaev},\ and\ \citenamefont {Nori}}]{bliokh2017optical_prl}%
  \BibitemOpen
  \bibfield  {author} {\bibinfo {author} {\bibfnamefont {K.~Y.}\ \bibnamefont
  {Bliokh}}, \bibinfo {author} {\bibfnamefont {A.~Y.}\ \bibnamefont
  {Bekshaev}},\ and\ \bibinfo {author} {\bibfnamefont {F.}~\bibnamefont
  {Nori}},\ }\bibfield  {title} {\bibinfo {title} {Optical momentum, spin, and
  angular momentum in dispersive media},\ }\href@noop {} {\bibfield  {journal}
  {\bibinfo  {journal} {Physical review letters}\ }\textbf {\bibinfo {volume}
  {119}},\ \bibinfo {pages} {073901} (\bibinfo {year}
  {2017}{\natexlab{a}})}\BibitemShut {NoStop}%
\bibitem [{\citenamefont {Bliokh}\ \emph
  {et~al.}(2017{\natexlab{b}})\citenamefont {Bliokh}, \citenamefont
  {Bekshaev},\ and\ \citenamefont {Nori}}]{bliokh2017optical_njp}%
  \BibitemOpen
  \bibfield  {author} {\bibinfo {author} {\bibfnamefont {K.~Y.}\ \bibnamefont
  {Bliokh}}, \bibinfo {author} {\bibfnamefont {A.~Y.}\ \bibnamefont
  {Bekshaev}},\ and\ \bibinfo {author} {\bibfnamefont {F.}~\bibnamefont
  {Nori}},\ }\bibfield  {title} {\bibinfo {title} {Optical momentum and angular
  momentum in complex media: from the abraham--minkowski debate to unusual
  properties of surface plasmon-polaritons},\ }\href@noop {} {\bibfield
  {journal} {\bibinfo  {journal} {New Journal of Physics}\ }\textbf {\bibinfo
  {volume} {19}},\ \bibinfo {pages} {123014} (\bibinfo {year}
  {2017}{\natexlab{b}})}\BibitemShut {NoStop}%
\bibitem [{\citenamefont {Uchida}\ \emph {et~al.}(2015)\citenamefont {Uchida},
  \citenamefont {Adachi}, \citenamefont {Kikuchi}, \citenamefont {Ito},
  \citenamefont {Qiu}, \citenamefont {Maekawa},\ and\ \citenamefont
  {Saitoh}}]{uchida2015generation}%
  \BibitemOpen
  \bibfield  {author} {\bibinfo {author} {\bibfnamefont {K.}~\bibnamefont
  {Uchida}}, \bibinfo {author} {\bibfnamefont {H.}~\bibnamefont {Adachi}},
  \bibinfo {author} {\bibfnamefont {D.}~\bibnamefont {Kikuchi}}, \bibinfo
  {author} {\bibfnamefont {S.}~\bibnamefont {Ito}}, \bibinfo {author}
  {\bibfnamefont {Z.}~\bibnamefont {Qiu}}, \bibinfo {author} {\bibfnamefont
  {S.}~\bibnamefont {Maekawa}},\ and\ \bibinfo {author} {\bibfnamefont
  {E.}~\bibnamefont {Saitoh}},\ }\bibfield  {title} {\bibinfo {title}
  {Generation of spin currents by surface plasmon resonance},\ }\href@noop {}
  {\bibfield  {journal} {\bibinfo  {journal} {Nature communications}\ }\textbf
  {\bibinfo {volume} {6}},\ \bibinfo {pages} {1} (\bibinfo {year}
  {2015})}\BibitemShut {NoStop}%
\bibitem [{Note1()}]{Note1}%
  \BibitemOpen
  \bibinfo {note} {Thorlabs, N-BK7 \protect \href
  {https://www.thorlabs.com/newgrouppage9.cfm?objectgroup_id=6973&tabname=N-BK7}{product
  details}}\BibitemShut {NoStop}%
\bibitem [{Note2()}]{Note2}%
  \BibitemOpen
  \bibinfo {note} {Schott, Optical glass \protect \href
  {https://www.schott.com/d/advanced_optics/ac85c64c-60a0-4113-a9df-23ee1be20428/1.17/schott-optical-glass-collection-datasheets-english-may-2019.pdf}{data
  sheets} May 2019}\BibitemShut {NoStop}%
\bibitem [{\citenamefont {Ordal}\ \emph {et~al.}(1985)\citenamefont {Ordal},
  \citenamefont {Bell}, \citenamefont {Alexander}, \citenamefont {Long},\ and\
  \citenamefont {Querry}}]{ordal1985optical}%
  \BibitemOpen
  \bibfield  {author} {\bibinfo {author} {\bibfnamefont {M.~A.}\ \bibnamefont
  {Ordal}}, \bibinfo {author} {\bibfnamefont {R.~J.}\ \bibnamefont {Bell}},
  \bibinfo {author} {\bibfnamefont {R.~W.}\ \bibnamefont {Alexander}}, \bibinfo
  {author} {\bibfnamefont {L.~L.}\ \bibnamefont {Long}},\ and\ \bibinfo
  {author} {\bibfnamefont {M.~R.}\ \bibnamefont {Querry}},\ }\bibfield  {title}
  {\bibinfo {title} {Optical properties of fourteen metals in the infrared and
  far infrared: Al, co, cu, au, fe, pb, mo, ni, pd, pt, ag, ti, v, and w.},\
  }\href@noop {} {\bibfield  {journal} {\bibinfo  {journal} {Applied optics}\
  }\textbf {\bibinfo {volume} {24}},\ \bibinfo {pages} {4493} (\bibinfo {year}
  {1985})}\BibitemShut {NoStop}%
\bibitem [{\citenamefont {KffiiLLOVA}\ \emph {et~al.}(1971)\citenamefont
  {KffiiLLOVA}, \citenamefont {Nomerovannaya},\ and\ \citenamefont
  {Noskov}}]{kffiillova1971optical}%
  \BibitemOpen
  \bibfield  {author} {\bibinfo {author} {\bibfnamefont {M.}~\bibnamefont
  {KffiiLLOVA}}, \bibinfo {author} {\bibfnamefont {L.}~\bibnamefont
  {Nomerovannaya}},\ and\ \bibinfo {author} {\bibfnamefont {M.}~\bibnamefont
  {Noskov}},\ }\bibfield  {title} {\bibinfo {title} {Optical properties of
  molybdenum single crystals},\ }\href@noop {} {\bibfield  {journal} {\bibinfo
  {journal} {Soviet Physics JETP}\ }\textbf {\bibinfo {volume} {33}} (\bibinfo
  {year} {1971})}\BibitemShut {NoStop}%
\bibitem [{\citenamefont {Malitson}(1965)}]{malitson1965interspecimen}%
  \BibitemOpen
  \bibfield  {author} {\bibinfo {author} {\bibfnamefont {I.~H.}\ \bibnamefont
  {Malitson}},\ }\bibfield  {title} {\bibinfo {title} {Interspecimen comparison
  of the refractive index of fused silica},\ }\href@noop {} {\bibfield
  {journal} {\bibinfo  {journal} {Josa}\ }\textbf {\bibinfo {volume} {55}},\
  \bibinfo {pages} {1205} (\bibinfo {year} {1965})}\BibitemShut {NoStop}%
\bibitem [{\citenamefont {Zeman}\ and\ \citenamefont
  {Schatz}(1987)}]{zeman1987accurate}%
  \BibitemOpen
  \bibfield  {author} {\bibinfo {author} {\bibfnamefont {E.~J.}\ \bibnamefont
  {Zeman}}\ and\ \bibinfo {author} {\bibfnamefont {G.~C.}\ \bibnamefont
  {Schatz}},\ }\bibfield  {title} {\bibinfo {title} {An accurate
  electromagnetic theory study of surface enhancement factors for silver, gold,
  copper, lithium, sodium, aluminum, gallium, indium, zinc, and cadmium},\
  }\href@noop {} {\bibfield  {journal} {\bibinfo  {journal} {Journal of
  Physical Chemistry}\ }\textbf {\bibinfo {volume} {91}},\ \bibinfo {pages}
  {634} (\bibinfo {year} {1987})}\BibitemShut {NoStop}%
\bibitem [{\citenamefont {Maier}(2007)}]{maier2007plasmonics}%
  \BibitemOpen
  \bibfield  {author} {\bibinfo {author} {\bibfnamefont {S.~A.}\ \bibnamefont
  {Maier}},\ }\href@noop {} {\emph {\bibinfo {title} {Plasmonics: fundamentals
  and applications}}}\ (\bibinfo  {publisher} {Springer Science \& Business
  Media},\ \bibinfo {address} {Berlin},\ \bibinfo {year} {2007})\BibitemShut
  {NoStop}%
\bibitem [{\citenamefont {Stephens}\ and\ \citenamefont
  {Malitson}(1952)}]{stephens1952index}%
  \BibitemOpen
  \bibfield  {author} {\bibinfo {author} {\bibfnamefont {R.~E.}\ \bibnamefont
  {Stephens}}\ and\ \bibinfo {author} {\bibfnamefont {I.~H.}\ \bibnamefont
  {Malitson}},\ }\bibfield  {title} {\bibinfo {title} {Index of refraction of
  magnesium oxide},\ }\href@noop {} {\bibfield  {journal} {\bibinfo  {journal}
  {J. Res. Natl. Bur. Stand.}\ }\textbf {\bibinfo {volume} {49}},\ \bibinfo
  {pages} {249} (\bibinfo {year} {1952})}\BibitemShut {NoStop}%
\bibitem [{\citenamefont {Pierce}\ and\ \citenamefont
  {Meier}(1976)}]{pierce1976photoemission}%
  \BibitemOpen
  \bibfield  {author} {\bibinfo {author} {\bibfnamefont {D.~T.}\ \bibnamefont
  {Pierce}}\ and\ \bibinfo {author} {\bibfnamefont {F.}~\bibnamefont {Meier}},\
  }\bibfield  {title} {\bibinfo {title} {Photoemission of spin-polarized
  electrons from gaas},\ }\href@noop {} {\bibfield  {journal} {\bibinfo
  {journal} {Physical Review B}\ }\textbf {\bibinfo {volume} {13}},\ \bibinfo
  {pages} {5484} (\bibinfo {year} {1976})}\BibitemShut {NoStop}%
\bibitem [{\citenamefont {Gao}\ \emph {et~al.}(2018)\citenamefont {Gao},
  \citenamefont {Xu}, \citenamefont {Gao}, \citenamefont {Zhang}, \citenamefont
  {Luo},\ and\ \citenamefont {Zhang}}]{gao2018surface}%
  \BibitemOpen
  \bibfield  {author} {\bibinfo {author} {\bibfnamefont {Z.}~\bibnamefont
  {Gao}}, \bibinfo {author} {\bibfnamefont {H.}~\bibnamefont {Xu}}, \bibinfo
  {author} {\bibfnamefont {F.}~\bibnamefont {Gao}}, \bibinfo {author}
  {\bibfnamefont {Y.}~\bibnamefont {Zhang}}, \bibinfo {author} {\bibfnamefont
  {Y.}~\bibnamefont {Luo}},\ and\ \bibinfo {author} {\bibfnamefont
  {B.}~\bibnamefont {Zhang}},\ }\bibfield  {title} {\bibinfo {title}
  {Surface-wave pulse routing around sharp right angles},\ }\href@noop {}
  {\bibfield  {journal} {\bibinfo  {journal} {Physical Review Applied}\
  }\textbf {\bibinfo {volume} {9}},\ \bibinfo {pages} {044019} (\bibinfo {year}
  {2018})}\BibitemShut {NoStop}%
\bibitem [{\citenamefont {Bozhevolnyi}\ \emph {et~al.}(2001)\citenamefont
  {Bozhevolnyi}, \citenamefont {Erland}, \citenamefont {Leosson}, \citenamefont
  {Skovgaard},\ and\ \citenamefont {Hvam}}]{bozhevolnyi2001waveguiding}%
  \BibitemOpen
  \bibfield  {author} {\bibinfo {author} {\bibfnamefont {S.~I.}\ \bibnamefont
  {Bozhevolnyi}}, \bibinfo {author} {\bibfnamefont {J.}~\bibnamefont {Erland}},
  \bibinfo {author} {\bibfnamefont {K.}~\bibnamefont {Leosson}}, \bibinfo
  {author} {\bibfnamefont {P.~M.}\ \bibnamefont {Skovgaard}},\ and\ \bibinfo
  {author} {\bibfnamefont {J.~M.}\ \bibnamefont {Hvam}},\ }\bibfield  {title}
  {\bibinfo {title} {Waveguiding in surface plasmon polariton band gap
  structures},\ }\href@noop {} {\bibfield  {journal} {\bibinfo  {journal}
  {Physical review letters}\ }\textbf {\bibinfo {volume} {86}},\ \bibinfo
  {pages} {3008} (\bibinfo {year} {2001})}\BibitemShut {NoStop}%
\bibitem [{\citenamefont {Stockman}(2004)}]{stockman2004nanofocusing}%
  \BibitemOpen
  \bibfield  {author} {\bibinfo {author} {\bibfnamefont {M.~I.}\ \bibnamefont
  {Stockman}},\ }\bibfield  {title} {\bibinfo {title} {Nanofocusing of optical
  energy in tapered plasmonic waveguides},\ }\href@noop {} {\bibfield
  {journal} {\bibinfo  {journal} {Physical review letters}\ }\textbf {\bibinfo
  {volume} {93}},\ \bibinfo {pages} {137404} (\bibinfo {year}
  {2004})}\BibitemShut {NoStop}%
\end{thebibliography}%
%

\end{document}